\documentclass[a4paper,11pt]{article}
\usepackage{pos}
\usepackage{tikz}
\usepackage{pgfplots}
\usepackage{slashed}
\usepackage{graphicx}
\usepackage{comment}
\usepackage{bm}

\usepackage{caption}
\usepackage{subcaption}
\DeclareMathOperator{\sign}{\rm sign}

\title{A Microscopic study of Magnetic monopoles in Topological Insulators}

\author{Shoto Aoki}
\author{Hidenori Fukaya}
\author*{Naoto Kan}
\author{Mikito Koshino}
\author{Yoshiyuki Matsuki}

\affiliation{Department of Physics, Osaka University,
Toyonaka, Osaka 560-0043, Japan}

\abstract{
In this article, we analyze a magnetic monopole in topological insulators.
The monopole obtain a fractional electric charge because of the Witten effect.
We consider this system with a microscopic view by adding the Wilson term to the ordinary Dirac Hamiltonian.
The Wilson term yields the positive mass shift to the effective mass of the electrons, then the curved domain-wall is dynamically generated around the monopole.
The zero-modes of the electrons are localized on the domain-wall, which can be identified as the source of the electric charge.
\par
Preprint number: OU-HET-1212
}

\FullConference{The 40th International Symposium on Lattice Field Theory (Lattice 2023)\\
July 31st - August 4th, 2023\\
Fermi National Accelerator Laboratory\\}

\begin{document}
\maketitle

\section{Introduction}
\label{sec:introduction}
A magnetic monopole has been intensively studied in particle physics.
In Maxwell theory, the monopole can be seen as a heavy point-like object with magnetic charge, i.e., the Dirac monopole \cite{Dirac1931-cv}, and the world line of the monopole is referred as the 't~Hooft loop.
In the modern sense, the 't~Hooft loop operator is the charged object of the one-form magnetic symmetry in Maxwell theory.
In grand unified theories, the monopole can be appeared as the 't~Hooft--Polyakov monopole which induces the proton decay \cite{Callan1982-ca, Rubakov1982-qs}.

We explore a behavior of the magnetic monopole in the so-called $\theta$ vacuum.
It is known that the monopole dresses electric charge in the non-zero $\theta$ vacuum, which is called the Witten effect \cite{Witten1979-fk}\footnote{See also related recent studies \cite{Pretko2017-gp,Yamamoto2020-gn,Abe2022-be,Hidaka2020-xv,Fukuda2020-lw,Sekine2020-ot,Agrawal2022-hu,Hamada2022-gq,Choi2022-xv,Yokokura2022-yf,Kan2023-km,Abe2023-ja}}.
The dressed electric charge, in general, becomes fractional.
The non-zero $\theta$ angle can be realized by introducing the axion field, but the following issues we study in this article are unclear in the effective field theoretical description: (1) what is the origin of the electric charge (it must be electrons, but unclear), (2) why the electron is bound to the monopole, (3) why the electric charge is fractional.

In this article, we focus on the $\theta=\pi$ vacuum, which is known that the vacuum describes the time-reversal symmetric topological insulator.
From the view point of a microscopic description\footnote{We find Refs.~\cite{Qi2008-wv, Sasaki2014-wp, Zirnstein2020-xb, Yamagishi1983-du} on the microscopic approach in condensed matter physics.}, we address the three issues mentioned above.
As a result of introducing the Wilson term to the Hamiltonian, we will find an interpretation of our issues.
Because of the positive correction to the effective mass from the Wilson term, the curved domain-wall\footnote{The curved domain-wall fermion is also studied in Refs.~\cite{Aoki2022-ly, Aoki2022-sc, Aoki2023-la}} is dynamically created around the monopole \cite{Aoki2023-hf}.
The chiral zero-modes of the electron are localized on the created domain-wall and bound to the monopole.
Due to the cobordism invariant nature of the Atiyah--Singer index, another domain-wall is required.
Since the 50\% of the wavefunction is located on the domain-wall around the monopole, we observe the fractional electric charge.

\section{A monopole in three-dimensions}
\label{sec:monopole}

\subsection{A naive Dirac equation with a monopole}
We start with solving the naive Dirac equation in the magnetic monopole background.
The Dirac Hamiltonian is 
	\begin{align}
		H=\gamma_0\left(\gamma^i\left(\partial_i-iA_i\right)+m\right),
	\end{align}
where the gamma matrices are defined by $\gamma_0=\sigma_3 \otimes {\bf 1}$ and $\gamma_i=\sigma_1\otimes \sigma_i$.
Let us put the Dirac monopole with magnetic charge $q_m$ at the origin.
Then the background gauge connection is given by
	\begin{align}
		A_1=-{q_m y \over r(r+z)}, \quad A_2={q_m x \over r(r+z)}, \quad A_3=0.
		\label{eq:monopole_config}
	\end{align}
Note that the magnetic charge is quantized, $q_m=\mathbb{Z}/2$, due to the Dirac quantization condition.
The Hamiltonian commutes with the total angular momentum $J_i$,
	\begin{align}
		J_i=L_i+{\sigma_i \over 2}, \quad L_i=-i\epsilon_{ijk} x_j\left(\partial_k-iA_k\right)-{q_m x_i \over r},
	\end{align}
with $[J_i,J_j]=i\epsilon_{ijk}J_k$, where $L_i$ is the orbital angular momentum.
As usual, we use ${\bf 1}\otimes J_3$ to parameterize the eigenstates. 
The square of the total angular momentum also commutes with the Hamiltonian: $[{\bf 1}\otimes J^2,H]=0$.
In addition, there is another operator which commutes with $H$ and ${\bf 1}\otimes J_i$.
We define the ``spherical'' operator as
	\begin{align}
		D^{S^2}=\sigma^i\left(L_i+{q_m x_i \over r}\right)+1.
	\end{align}
We will see a physical meaning of this operator later.
We can easily check $[\sigma_3\otimes D^{S^2},H]=0$ and $[J_i,D^{S^2}]=0$.

Let us solve the Dirac equation, $H\psi=E\psi$.
We find the normalizable zero-mode ($E=0$) solution localized at the origin with $j=|q_m|-1/2$:
	\begin{align}
		\psi_{j,j_3,0}(r,\theta,\phi)={C \over r}e^{-|m|r}
			\begin{pmatrix}
				1 \\
				\sign(m)\sign(q_m)
			\end{pmatrix}
		\otimes \chi_{j,j_3,0}(\theta,\phi),
	\label{eq:naive_sol}
	\end{align}
where we introduced the eigenstate of $D^{S^2}$ satisfying $D^{S^2}\chi_{j,j_3,0}=0$.
The solution is bound to monopole, but it still not enough to explain the following issues: 
The solution Eq.~\eqref{eq:naive_sol} can exist for both positive and negative mass.
The Witten effect predicts that the monopole becomes dyon in the topological insulator with $m<0$, but not in the normal insulator with $m>0$.
We cannot describe this prediction in the microscopic sense unless we impose the chiral boundary condition at the origin by hand.
It also remains unclear why the electric charge becomes fractional.
In Ref.~\cite{Yamagishi1983-du}, summing all charges within the Dirac sea and comparing it to the charge configuration in the absence of a monopole yields half of an electric charge, yet this can only be achieved following a regularization process that disrupts charge conservation.

\subsection{A regularized Dirac equation}
To explore the issues we saw at the end of the previous section, let us define the regularized Dirac Hamiltonian,
	\begin{align}
		H_{\rm reg}=H+\gamma_0{D_i^\dagger D^i \over M_{\rm PV}}=\gamma_0\left(\gamma^iD_i+m+{D_i^\dagger D^i \over M_{\rm PV}}\right).
	\end{align}
The physical meaning of the new term we added is as follows.
In continuum field theory, to obtain finite results, we must regularize the theory.
We now regularize our theory using the Pauli--Villars (PV) regularization.
Then the partition function is given by
	\begin{align}
		Z=\det \left({\slashed{D}+m \over \slashed{D}+M_{\rm PV}}\right)	=\det \left[{1 \over M_{\rm PV}}\left(\slashed{D}+m+{D^\dagger_\mu D^\mu \over M_{PV}}+{\cal O}(1/M_{\rm PV}^2,m/M_{\rm PV},F_{\mu\nu}/M_{\rm PV})\right)\right],
	\end{align}
where $M_{\rm PV}$ is the mass of the PV field \footnote{Precisely, in order to completely regularize the theory, we have to introduce multiple PV fields. However, a single PV field is sufficient for our current purpose.}.
Thus we can interpret the term we added to the Hamiltonian as the leading term of the large $M_{\rm PV}$ expansion.
We refer the additional term as the Wilson term since the Laplacian corresponds to the Wilson term in lattice gauge theory.

Before finding the zero-mode eigenstates of the regularized Hamiltonian $H_{\rm reg}$, let us see the role of the Wilson term schematically.
Since the Laplacian $D_i^\dagger D^i$ is alway positive, the mass shift due to the Wilson term is also always positive when we take $M_{\rm PV}$ positive.
In the topological insulator with $m<0$, the Wilson term can change the sign of the effective mass,
	\begin{align}
		m<0\quad \to \quad m_{\rm eff}=m+{D_i^\dagger D^i \over M_{\rm PV}}\sim m+{1 \over M_{\rm PV}r_1^2}>0,
	\end{align}
when the magnetic flux is concentrated in the region $r<r_1$.
It implies that the inner region $r<r_1$ becomes a normal insulator, and the spherical domain-wall is dynamically created and the chiral edge-modes appear on it.
Note that this does not happen in the normal insulator with $m>0$ since we have no sign flip of the mass around the monopole.

\begin{figure}[h]
		\centering 
		\includegraphics[scale=0.5]{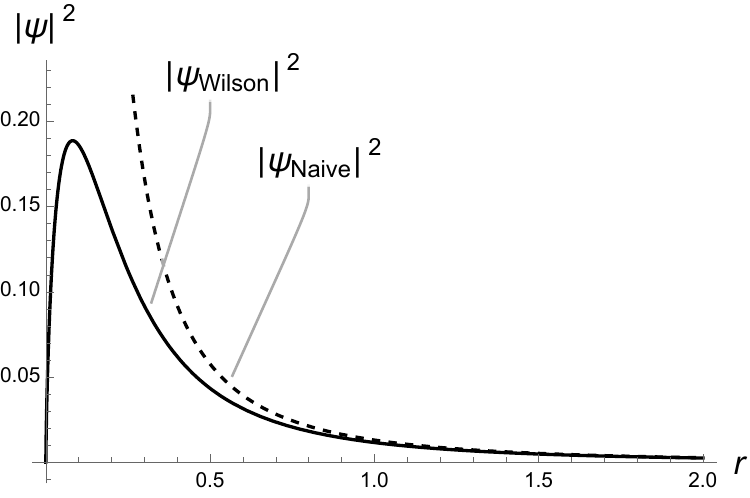}
		\caption{\label{fig:wavefunctions} The plot of the zero-mode solution with the Wilson term in Eq.~\eqref{eq:reg_sol} and without the Wilson term in Eq.~\eqref{eq:naive_sol}. We set $q_m=1/2$, $m=-0.1$ and $M_{\rm PV}=10$. Figure from Ref.~\cite{Aoki2023-hf}.}
	\end{figure}

Let us investigate the regularized Dirac Hamiltonian with the monopole background Eq.~\eqref{eq:monopole_config}.
The rotational symmetry is still genuine, $[H_{\rm reg},{\bf 1}\otimes J_i]$, but the spherical operator does not commute with the Laplacian: $[D^{S^2},D_i^\dagger D^i]\neq0$.
We find solution of the zero-mode for $r_1\to 0$,
	\begin{align}
		\psi_{j,j_3}^{\rm mono}(r,\theta,\phi)={B e^{-M_{\rm PV}r/2} \over \sqrt{r}}I_\nu(\kappa r)
			\begin{pmatrix}
				1 \\
				-\sign(q_m)
			\end{pmatrix}
		\otimes \chi_{j,j_3,0}(\theta,\phi),	
	\label{eq:reg_sol}
	\end{align}
where $\nu=\sqrt{4|q_m|+1}/2$, $\kappa=M_{\rm PV}\sqrt{1+4m/M_{\rm PV}}/2$, and $I_\nu(z)$ is the modified Bessel function of the first kind.
We note that since $I_\nu(\kappa r)\sim \exp(M_{\rm PV}r/2-|m|r)/\sqrt{\pi M_{\rm PV}r}$ in the $M_{\rm PV}\to \infty $ limit, the solution coincides with the naive solution Eq.~\eqref{eq:naive_sol}.
We also note that the solution has a peak at $r=|q_m|/M_{\rm PV}$ only when the mass $m$ is negative.
As is expected, this implies that the the domain-wall is created by the Wilson term.
We show the the zero-mode solution with and without the Wilson term in Fig.~\ref{fig:wavefunctions}.
We have set the parameters with $q_m=1/2$, $m=-0.1$ and $M_{\rm PV}=10$.

\subsection{Half-integral charge}

So far, we implicitly considered a $\mathbb{R}^3$ space, but to discuss nature of topology, we need to compactify the space on a $S^3$.
Then topological insulator region with $m_{\rm eff}<0$ seems to have the topology of a disk $D^3$ with a small $S^2$ boundary at $r =r_1$.
However, due to the cobordism invariance of the Atiyah--Singer index, the index must be zero:
	\begin{align}
	{\rm Ind}\, D^{S^2}={1 \over 4\pi} \int_{S^2}F={1 \over4\pi}\int_{D^3}dF=0,
	\end{align}
which contradicts with ${\rm Ind}\, D^{S^2}=2q_m$ if there are the only chiral zero modes.
A resolution is to create another domain-wall at, say, $r=r_0$, outside of the topological insulator.
Another zero-modes are localized at the outside domain-wall, and the index is kept trivial:
	\begin{align}
		\int_M dF=\int_{S^2_{\rm mono}\cup S^2_{\rm out}}F=0,
	\end{align}
where $M$ is the region of the topological insulator, $S^2_{\rm mono}$ and $S^2_{\rm out}$ correspond to the domain-wall around the monopole and outside domain-wall, respectively.
The zero modes localized on $S^2_{\rm mono}$ and on $S^2_{\rm out}$ are mixed by the tunneling effect for finite $r_0$, i.e., $\psi=\alpha \psi^{\rm mono}_{j,j_3}+\beta \psi^{\rm DW}_{j,j_3}$, where $\alpha$ and $\beta$ are constant coefficients which parameterize the mixing, $\psi^{\rm mono/DW}_{j,j_3}$ are the localized zero modes around the monopole/outside domain-wall.
We can show that $(\psi^{\rm mono}_{j,j_3})^\dagger H \psi^{\rm mono}_{j,j_3}=(\psi^{\rm DW}_{j,j_3})^\dagger H \psi^{\rm DW}_{j,j_3}=0$ for the diagonal parts and $(\psi^{\rm mono}_{j,j_3})^\dagger H \psi^{\rm DW}_{j,j_3}=(\psi^{\rm DW}_{j,j_3})^\dagger H \psi^{\rm mono}_{j,j_3}=: \Delta \in \mathbb{R}$ for the off-diagonal perts.
Then we can obtain that $\alpha=\pm \beta$ and the split energy is $E=\pm \Delta$.
As a result, the 50\% of the zero mode state is located at the monopole, while the other is sit at the outside domain-wall, thus the dressed electric charge of the monopole becomes half-integer.

\section{Numerical results}
\label{sec:numerical}

\subsection{Lattice setup}
We prepare the three-dimensional hyper-cubic lattice with size $L=31$\footnote{In Ref.~\cite{Aoki2023-hf}, we studied the three cases with $L=23, 31$ and $47$.}.
Imposing on open boundary conditions, we put a monopole at $\bm{x}=\bm{x}_{\rm m}=(L/2,L/2,L/2)$ and anti-monopole at $\bm{x}=\bm{x}_{\rm a}=(L/2,L/2,1/2)$.
The continuum vector potential at $\bm{x}=(x,y,z)$ is given by
	\begin{align}
		A_1(\bm{x})&=q_m\left[{-(y-y_{\rm m}) \over |\bm{x}-\bm{x}_{\rm m}|(|\bm{x}-\bm{x}_{\rm m}|+(z-z_{\rm m}))}-{-(y-y_{\rm a}) \over |\bm{x}-\bm{x}_{\rm a}|(|\bm{x}-\bm{x}_{\rm a}|+(z-z_{\rm a}))}\right], \\
		A_2(\bm{x})&=q_m\left[{x-x_{\rm m} \over |\bm{x}-\bm{x}_{\rm m}|(|\bm{x}-\bm{x}_{\rm m}|+(z-z_{\rm m}))}-{x-x_{\rm a} \over |\bm{x}-\bm{x}_{\rm a}|(|\bm{x}-\bm{x}_{\rm a}|+(z-z_{\rm a}))}\right], \\
		A_3(\bm{x})&=0,
	\end{align}
with $q_m\in \mathbb{Z}/2$.
We assume a position-dependent mass $m(\bm{x})$ for fermions.
We set negative mass $m(\bm{x})=-m_0$ for $r=\sqrt{|\bm{x}-\bm{x}_{\rm m}|}<r_0=3L/8$, while positive mass $m(\bm{x})=+m_0$ for $r>r_0$ with $m_0(L+1)=14$.
The Wilson Dirac Hamiltonian is given by
	\begin{align}
		H_{\rm Wilson}=\gamma^0\left[\sum^3_{i=1}\left(\gamma^i{\nabla^f_i+\nabla^b_i \over 2}-{1 \over 2}\nabla_i^f\nabla^b_i\right)+m(\bm{x})\right],
	\end{align}
where the forward covariant difference is $\nabla_i^f\psi(\bm{x})=U_i(\bm{x})\psi(\bm{x}+\bm{e}_i)-\psi(\bm{x})$, and the backward difference is $\nabla^b_i\psi(\bm{x})=\psi(\bm{x})-U^\dagger_i(\bm{x}-\bm{e}_i)\psi(\bm{x}-\bm{e}_i)$.
The link variable $U_j(\bm{x})$ is defined as
	\begin{align}
		U_j(\bm{x})=\exp\left(i\int^1_0 A_j(\bm{x}+\bm{e}_jl)dl\right).
	\end{align}

\subsection{Numerical analises}
\begin{figure}[h]
	\centering 
	\includegraphics[scale=0.4]{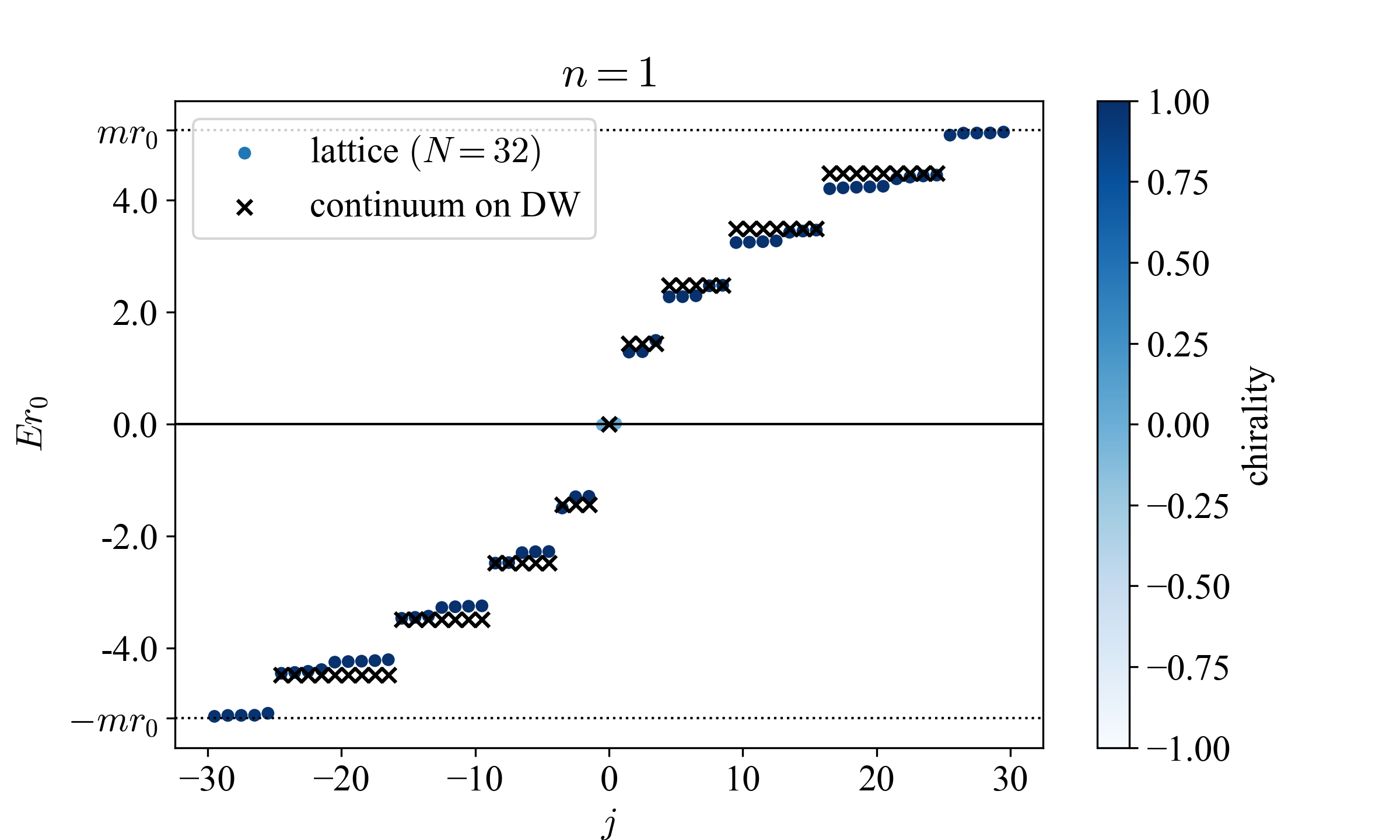}
	\caption{\label{fig:eigen_modes} The spectrum of the energy eigenvalue $E_k$ with magnetic charge $q_m=1/2$. Figure from Ref.~\cite{Aoki2023-hf}}
\end{figure}

We compute the energy eigenvalues and eigenfunctions of $H_{\rm Wilson}$ with $q_m=1/2$ and  as $E_k$ and $\phi_k(\bm{x})$, respectively.
We plot the eigenvalues $E_k$ in units of $r_0$ in Fig.~\ref{fig:eigen_modes}.
We show the results of the continuum calculation without the Wilson term by cross symbols.
We also show the chirality expectation value measured by
	\begin{align}
		\sum_{\bm{x}}\phi^\dagger_k(\bm{x})\left(\sigma_1\otimes \sigma_r\right)\phi_k(\bm{x})
	\end{align}
using the color gradation.
We find the two nearest zero-modes.

\begin{figure}[h]
  \begin{minipage}[b]{0.45\linewidth}
    \centering
    \includegraphics[keepaspectratio, scale=0.44]{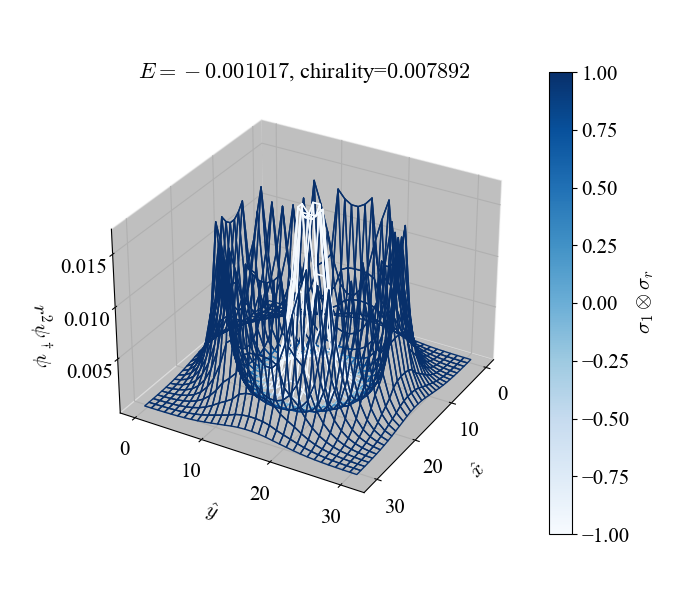}
  \end{minipage}
  \begin{minipage}[b]{0.45\linewidth}
    \centering
    \includegraphics[keepaspectratio, scale=0.44]{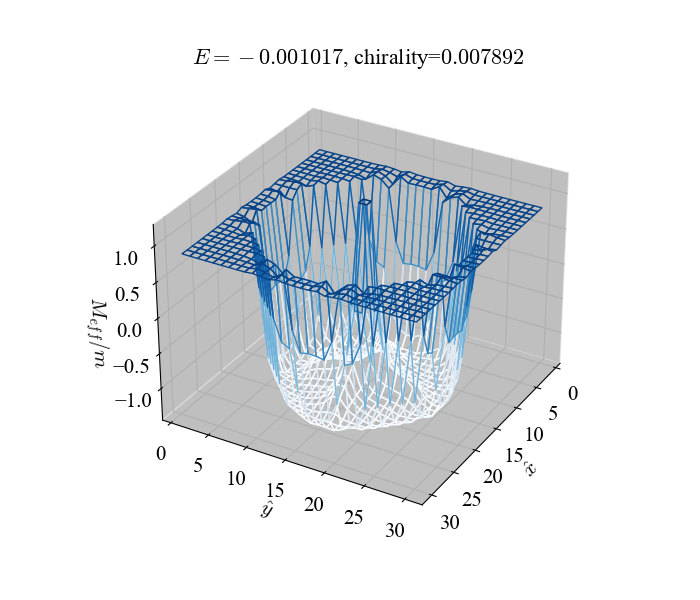}
  \end{minipage}
  \caption{\label{fig:amplitude_mass} Left panel: the amplitude of the nearest zero-mode $A_1(\bm{x})$ at $z=16$ with $q_m=1/2$. Right panel: the local effective mass of the nearest zero-mode $m_{\rm eff}(\bm{x})$. Figure from Ref.~\cite{Aoki2023-hf}}
\end{figure}

In the left panel of Fig.~\ref{fig:amplitude_mass}, we plot the local amplitude of the (nearest) zero-mode,
	\begin{align}
		A_1(\bm{x})=\phi_1(\bm{x})^\dagger \phi_1(\bm{x})r^2
	\end{align}
normalized by $r^2$ at the $z=16$ slice.
The amplitude has two peaks around $r=0$ and $r=r_0$.
The 50\% of the state is located around the monopole, while the other 50\% is located at $r=r_0$, which implies that we observe the half electric charge.

In the right of Fig.~\ref{fig:amplitude_mass}, the distribution of the local effective mass
	\begin{align}
		m_{\rm eff}(\bm{x})=\phi_1(\bm{x})^\dagger \left(-{1 \over 2}\sum_i \nabla^f_i\nabla^b_i+ m(\bm{x})\right)\phi_1(\bm{x})r^2/A_1(\bm{x})
	\end{align}
at the $z=16$ slice is plotted, as well.
We see that the small island of the normal insulator, i.e., the positive mass region appears around the monopole; the domain-wall is dynamically created.

\section{Summary}
In this article, we have microscopically studied the magnetic monopole in the topological insulators.
In the analysis of the naive Dirac equation, it is possible to find the zero mode solution localized around a monopole. 
However, the solution cannot satisfactorily explain why the Witten effect occurs only when the fermion mass is negative, and why the electric charge dressed by the monopole becomes fractional.
To explain this, we have considered a regularized Hamiltonian with the Wilson term. Due to the effects of the Wilson term, the domain wall is created around the monopole only when the fermion mass is negative. 
As a result, the zero modes localize on the domain wall.
Furthermore, the cobordism invariance of the Atiyah--Singer index theorem requires the existence of the outside domain wall, and we have concluded that the $50\%$ of the wavefunction of the zero modes are localized around the monopole, so that the electric charge becomes fractional.
we have also numerically analyzed the creation of the domain-wall around the monopole and the emergence of the chiral zero modes.

\subsection*{Acknowledgements}
We thank Mikio~Furuta, Masahiro~Hotta, Takuto~Kawakami, Tsunehide~Kuroki, Okuto~Morikawa, Tetsuya~Onogi, Shigeki~Sugimoto, Kazuki~Yamamoto, Masahito~Yamazaki, and Ryo~Yokokura for useful discussions. 
The work of SA was supported by JSPS KAKENHI Grant Number JP23KJ1459. The work of HF and NK was supported by JSPS KAKENHI Grant Number JP22H01219. The work of MK was supported by JSPS KAKENHI Grant Number JP21H05236, JP21H05232, JP20H01840, and JP20H00127, and by JST CREST Grant Number JPMJCR20T3, Japan.
\bibliographystyle{JHEP}
\bibliography{ref}

\end{document}